\documentclass[review]{elsarticle}

\usepackage{amssymb}
\usepackage{amssymb}
\usepackage{graphicx}
\usepackage{amsmath}
\usepackage{lineno,hyperref}
\modulolinenumbers[5]
\usepackage{algorithm}
\usepackage{algorithmic}
\usepackage{mathtools}
\usepackage{amssymb}
\usepackage{amsmath}
\usepackage{graphicx}
\usepackage{color}
\newcommand\numberthis{\addtocounter{equation}{1}\tag{\theequation}}
\usepackage{lipsum}
\usepackage{mathtools}
\usepackage{cuted}

\journal{Physical Communication}

\usepackage
[
    ,letterpaper
    ,left       = 1in
    ,right      = 1in
    ,top        = 1.5in
    ,headheight = 1in
    ,headsep    = .3in
]{geometry}

\journal{Physical Communication}

\begin{document}

\begin{frontmatter}

\title{Non-Orthogonal Multiple Access with Spatial Modulation in Downlink Coordinated Multipoint Transmission}

\author{Denny Kusuma Hendraningrat$^\bullet$}
\ead{dennykh@ieee.org}
\author{Bhaskara Narottama$^\dagger$}
\ead{bhaskara@kumoh.ac.kr}
\author{Soo Young Shin$^\star$} \corref{cor1}
\ead{wdragon@kumoh.ac.kr}

\address{ Department of IT Convergence Engineering \\ Kumoh National Institute of Technology, South Korea}
\tnotetext[t1]{Wireless and Emerging Networks System (WENS) Lab., Dept. of IT Convergence Engineering, Kumoh National Institute of Technology (KIT), 39177, Gumi, South
Korea.}
\cortext[cor1]{Corresponding Author}

\begin{abstract}
In this paper, a joint transmission coordinated multi-point based non-orthogonal multiple access (JT-CoMP NOMA) combined with spatial modulation (SM), termed as JT-CoMP NOMA-SM, is proposed to enhance capacity. User capacity and ergodic sum capacity (ESC) of $M$ number coordinated multi-point base stations (CoMP BSs) within $N$ number of cells are analyzed by considering imperfect successive interference cancellation (SIC) and imperfect channel state information (CSI). The performances of the proposed system are compared with non-orthogonal multiple access (NOMA), and joint transmission coordinated multi-point combined with virtual user pairing based non-orthogonal multiple access (JT-CoMP VP-NOMA) by both simulation and analysis. The results show that the proposed system has the same cell center user (CCU) capacity compared to JT-CoMP VP-NOMA and a higher cell edge user (CEU) capacity than the other schemes. ESC of the proposed system outperforms the other schemes due to enhancing CEU capacity. Imperfect SIC and imperfect CSI may degrade capacity. %Average ESC of the proposed system decreases if the number of CoMP BSs is increased.
The proposed system can maintain CEU capacity better than the other schemes if the number of cells is increased. It happens because SM works beyond Shannon upper bounds which can mitigate inter-cell interference (ICI).
\end{abstract}


\end{frontmatter}

\section{Introduction}\label{sec1}
%\linenumbers
%In recent years, the number of devices connected to the communication networks has been increasing significantly \cite{elkhodr_emerging}. To provide quality of service (QoS) for the massive connectivity, there is an inevitable need for new mobile communications systems to enhance its capabilities \cite{zaidi_QoS}. 
Non-orthogonal multiple access (NOMA) is a multiple access technique considered for further improving the spectral efficiency compared to orthogonal multiple access (OMA) [1-4]. NOMA supports multiple users multiplexed over a particular resource by allocating lower power to cell center user (CCU) than cell edge user (CEU). In addition, CCU uses successive interference cancellation (SIC) to decode own signal by canceling CEU signal, while CEU performs direct decoding without any SIC [3,4]. 
\par
In downlink multi-cell network, inter-cell interference (ICI) is one of the critical issues because it may degrade the system performance, such as user throughput and cell capacity. Coordinated multi-point (CoMP) is introduced as an ICI mitigation technique that may escalate capacity of NOMA [5-11]. In a conventional CoMP, one CEU is served by joint transmissions of multiple coordinated BSs simultaneously to improve its performance [5-7]. In [7], it exploited conventional CoMP based NOMA scheme by considering both imperfect channel state information (CSI) and imperfect SIC simulated in multi-cell scenarios.  However, conventional CoMP has a limitation that only can handle one CEU within $M$ number of coordinated multi-point base stations (CoMP BSs). 
\par
To handle multiple CEUs, several multiple CoMP-user techniques have been developed [8-11]. 
%In \cite{Eryani_generalized}, a novel generalized COMP-enabled NOMA (GCoMP-enabled NOMA) was further exploited to handle multiple CEUs based on conventional CoMP technique. It classified users into clusters at a certain frequency sub-band to be served by using virtual multiple input multiple output (MIMO) antenna. By assuming the number of BS and MIMO antenna are fixed, then inter-cluster interference will be high when the cluster size is large. 
In [11], a joint transmission coordinated multi-point combined with virtual user pairing based non-orthogonal multiple access (JT-CoMP VP-NOMA) was proposed to handle multiple CEUs paired with one CCU. Evaluating JT-CoMP VP-NOMA in [11], it assumed a cluster contains a number of both cells and users have been predetermined. JT-CoMP VP-NOMA may not be efficient while the cluster contains a lot of cells and/or each cell contains a lot of CEUs [12]. Considering each cell contains two users, it may degrade CEU capacity because both bandwidth and BS power are divided by one CCU and $M$ number of CEUs, although each CEU receives signals from $M$ number of CoMP BSs. However, the distribution of users in a cell is random in practice; the number of CCU is equal to CEU or the number of CCU may be more/less than CEU.
\begin{figure*}[!b]
\centering
  \includegraphics[width=16cm,height=9.7cm]{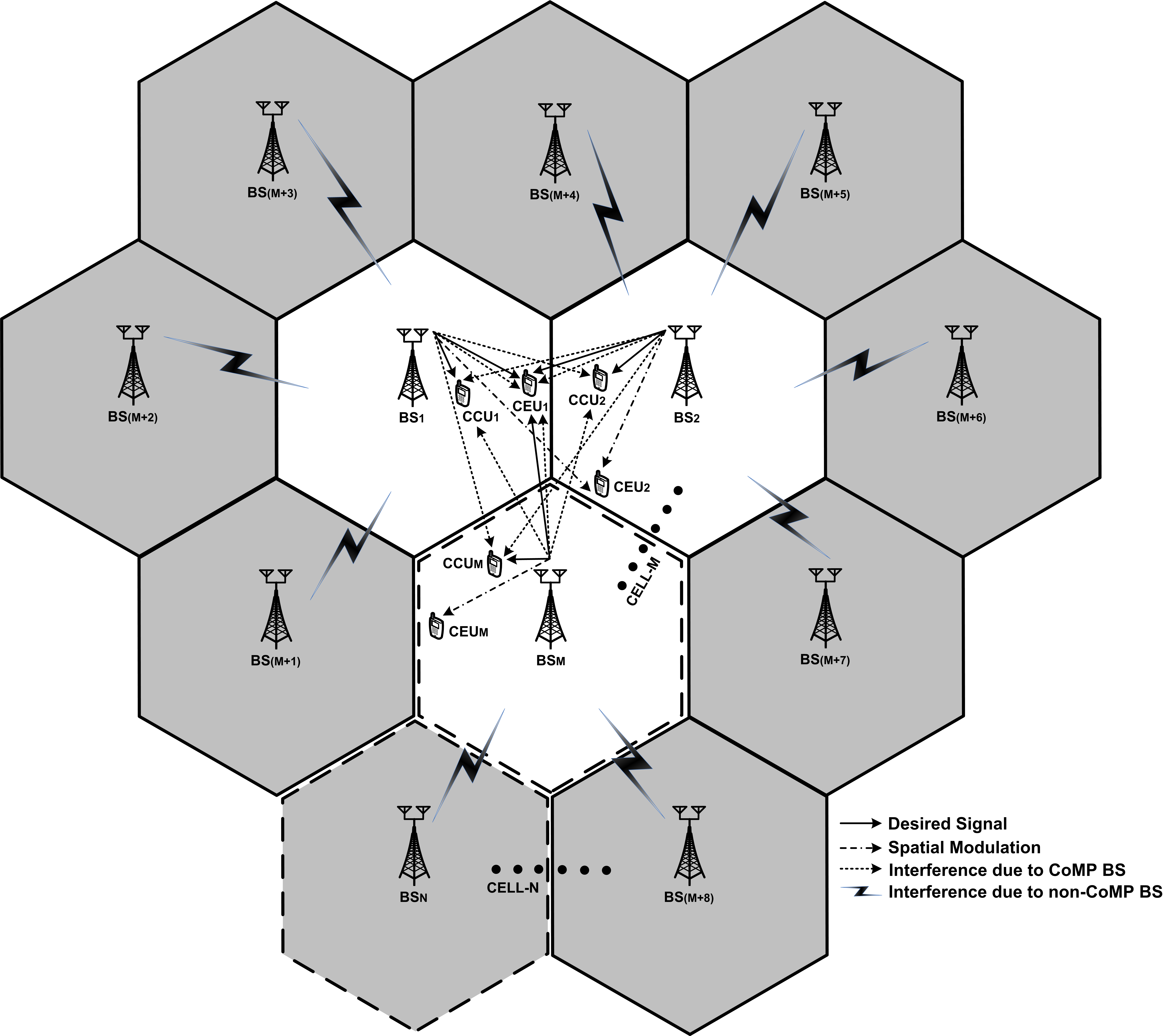}
    \caption {System model of the proposed JT-CoMP NOMA-SM.}
    \label{Fig1}
\end{figure*}
% 
%Furthermore, combining NOMA and SM should be able to increase the user performance \cite{kim_performance,kim_selective}.
%{This paper is motivated by previous research in \cite{shin_comp}, which is evaluated related to cell edge performance. It proposed JT-CoMP VP-NOMA to handle multiple CEUs in order to enhance ESC simulated in multi-cell scenarios. }

%Using JT-CoMP  \cite{murti_exploiting} combined with VP-NOMA \cite{shahab_virtual,shahab_performance,kader_non-orthogonal}, term as JT-CoMP VP-NOMA, to handle multiple CEUs simulated in multi-cell scenarios is not efficient because ESC increases by enhancing CCU capacity while the CCU may recognize CEUs signal and cancel it using SIC. 
%On the other hand, SM represents family of modulation schemes, which transmit additional information using antennas as its index. SM, which indices of the antenna blocks of the considered communications systems, conveys additional information bits beyond Shannon upper bounds \cite{soujeri_advanced,mesleh_spatial}. SM can be utilized to handle generalized random user distribution, which only increases the number of BS antenna while the number of user in a cell is increased. Furthermore, combination of NOMA and SM should be able to increase the user performance \cite{kim_performance,kim_selective}.
%to enhance efficiency of the power allocated to the CEU, % to handle other CEUs which cannot be served by using conventional JT-CoMP NOMA
Regarding the issues, spatial modulation (SM) technique is considered may handle multiple CCUs and/or multiple CEUs efficiently [13-17]. Considering conventional CoMP can handle $M$ number of CCUs and one CEU within $M$ number of CoMP BSs, SM is utilized to handle the other users that cannot be served by using conventional CoMP [16,17]. In this paper, conventional CoMP is used because power allocated for CEU can be optimized while the number of CoMP BSs is increased. By combining conventional CoMP and SM, it is expected may enhance user capacity. An opportunity for further improve user capacity appears, by coordinating SM antennas to transmit a specific user, because CoMP and SM techniques are combined simultaneously.

%In \cite{kim_selective}, it analyzed non-orthogonal multiple access space shift keying (NOMA-SSK) to enhance spectral efficiency by multiplexing the CEU in both frequency and spatial domains.

The main contributions of this paper are listed as follows: 
\begin{itemize} 
\item[$\bullet$] This paper proposes JT-COMP NOMA-SM to handle random user distribution based on conventional CoMP. It is expected may improve user capacity by mitigating ICI compared to both NOMA and JT-COMP VP-NOMA.
\item[$\bullet$] By using CoMP technique, this paper proposes coordinated SM (CoSM) for further improve user performance.
\item[$\bullet$] This paper conducts a mathematical analysis of the proposed system.\end{itemize}

The rest of this paper is organized as follows: Section \ref{sec_sys} presents system model of the proposed JT-CoMP NOMA-SM. Section \ref{sec_erg} conducts a closed-form of ergodic sum capacity (ESC) for the proposed system. Then, section \ref{sec_result} provides simulation result and analysis. Finally, section \ref{sec_conclusion} concludes the overall of this paper.

\section{System Model} \label{sec_sys}
%In this paper, we have $N$-cells wrapped around system which each cell contains one BS, one CCU, and one CEU. 
This paper considers $M$ number of CoMP BSs within $N$ number of cells. To keep in line with previous work for comparison, each cell which is coordinated contains one BS, one CCU, and one CEU [11]. Each $i^{th}$ cell represents coverage of each $i^{th}$ BS $(i\in 1,2,...,N)$. Considering $N$ number of cells, $M$ number of cells $(2 \leq M \leq N$ and $M \subset  N)$ are coordinated while the other cells do not apply coordination. Considering $M$ number of cells, the $j^{th}$
CCU is represented with $CCU_j$ $(j\in 1,2,...,M)$, whereas the $k^{th}$ CEU is expressed with $CEU_k$ $(k\in 1,2,...,M)$. In this paper, user positions are generated based on the random user algorithm. Furthermore, $r_{ij}$ and $r_{ik}$ represent the $j^{th}$ CCU and the $k^{th}$ CEU distances from the $i^{th}$ BS ground, respectively. The maximum cell radius from the BS ground is normalized by $R=1$. In this case, BS antenna high and BS distances to the other BSs are assigned by $0.01R$ and $2R$, respectively. Moreover, $d_{ij}$ and $d_{ik}$ can be calculated, as the $j^{th}$ CCU and  the $k^{th}$ CEU distance from the $i^{th}$ BS antenna respectively, using the concept of trigonometry [7,11].
%
%\begin{figure}[!t]
%    \centering
%    \includegraphics %[height=9cm,width=8.5cm]{Fig2.png} \\
%    \caption {Considering scenarios: %(a) OMA, (b) NOMA, \\ (c) JT-CoMP %VP-NOMA, and (d) JT-CoMP NOMA-SM.}
%    \label{Fig2}
%    \end{figure}
% 

Considering $M$ number of CoMP BSs, the proposed JT-CoMP NOMA-SM can be divided into JT-CoMP NOMA and SM systems as shown in Fig. \ref{Fig1}. In JT-CoMP NOMA system, $M$ number of CCUs ($CCU_1$, $CCU_2$,...,$CCU_M$) and a CEU are assigned as NOMA users. In this paper, the first CEU ($CEU_1$) is assigned as a CoMP user, while all CCUs do not apply CoMP technique. In SM system, other CEUs ($CEU_2,...,CEU_M$), which are not included in the JT CoMP NOMA system, are assigned as SM users.
%SM delivers information to the the SM users by using antenna index.

\subsection{JT-CoMP NOMA system}
In this system, a conventional JT-CoMP NOMA is applied for $M$ number of CoMP BSs within $N$ number of cells. This paper assumes CoMP user, i.e. $CEU_k$ for $k=1$, is transmitted by multiple coordinated BSs simultaneously. It paired with non-CoMP users, i.e. $CCU_j$ for $1 \leq j \leq M$, in each cell. 
\par
In this paper, each $j^{th}$ CCU is allocated the same resource as CoMP user, where total bandwidth of each BS is normalized by $B=1$. Total power of each BS is normalized by $P=1$. Power allocation factor for CCU and CEU are assigned by $\alpha$ and $\beta$, respectively. This paper assumes $\alpha=0.1$, while $\beta$ can be calculated with 1-$\alpha$. Then, $P_j=\alpha P$ and $P_{k}$=$\beta P$ are assigned as power allocation for CCU and CoMP user respectively, where $k=1$ is assigned for each $k$ symbol mentioned in JT-CoMP NOMA system. %For simplicity, each BS is assumed to have two antennas $(M_{Ti}=2)$. $M_{Ti}$ represents the number of the $i^{th}$ BS antenna. 
%Furthermore, a downlink $N$ number of cells model is adopted from the previous works \cite{murti_exploiting,denny_virtual}. It defined $r_{ij}$ and $r_{ik}$ as the $j^{th}$ CCU and the $k^{th}$ CEU distances from the $i^{th}$ BS ground, respectively. The maximum cell radius from the BS ground is normalized by $R=1$. In this case, BS antenna high and BS distances to its neighboring BS are assigned by $0.01R$ and $2R$, respectively. Moreover, $d_{ij}$ and $d_{ik}$ can be calculated, as the $j^{th}$ CCU and  the $k^{th}$ CEU distance from the $i^{th}$ BS antenna respectively, using the concept of trigonometry.

In this paper, imperfect SIC and imperfect CSI are considered in the simulation parameters as a representation of real condition [7,11]. For simplicity, links between each antenna in a BS to a user are assumed to have the same channel characteristic. Imperfect CSI is modeled with channel estimation error, where a priory of a variance of the error estimation is known. The channel estimation error for the link between the $i^{th}$ BS to the $j^{th}$ CCU can be modeled by $h_{\varepsilon ij}$=$h_{ij}$-${\hat{h}}_{ij}$ [7,11]. In addition, channel estimation error for the link between the $i^{th}$ BS to CoMP user can be written as $h_{\varepsilon ik}$=$h_{ik}$-${\hat{h}}_{ik}$. It is assumed channel over each link is independent Rayleigh flat fading with channel coefficients $h_{\varepsilon ij}\sim CN(0,\sigma_{\varepsilon ij})$, and $h_{\varepsilon ik}\sim CN(0,\sigma_{\varepsilon ik})$. Furthermore, received signals for the $j^{th}$ CCU and CoMP user need consider a channel estimation gain $|\hat{h}_{ij}|^2$ and $|\hat{h}_{ik}|^2$, respectively.  Therefore, the channel estimation characteristic for the $j^{th}$ CCU and CoMP user are assumed to be distributed independently with mean zero, which can be modeled as $\hat{h}_{ij}\sim CN(0,\hat{\sigma}_{ij}=d_{ij}^{-v}-\sigma_{\varepsilon ij})$, and $\hat{h}_{ik}\sim CN(0,\hat{\sigma}_{ik}=d_{ik}^{-v}-\sigma_{\varepsilon ik})$, respectively. In this case, $\hat{\sigma}_{ij}$ ̂and $\hat{\sigma}_{ik}$ represent estimation variance for the link from the $i^{th}$ BS antenna to the $j^{th}$ CCU and CoMP user, respectively, where $v$ represents the path-loss exponent. For simplicity, the variance of each channel estimation error is assumed fixed.

%Since perfect knowledge of SIC in $CCU_j$ cannot decode own message perfectly and a channel is not always available in practice, so imperfect SIC \cite{shahab_user} and imperfect CSI \cite{cheng_NOMA,wang_impact} need to be considered. 

\subsection{SM system}
Considering $M$ number of CoMP BSs, conventional CoMP already applied to all CCUs ($CCU_1$,$CCU_2$,...,\\$CCU_M$) and one CEU ($CEU_1$) in this case, then the other users ($CEU_2$,...,$CEU_M$) are not allocated bandwidth by BS. Furthermore, SM users, i.e. $CEU_k$ for $2 \leq k \leq M$, are handled by using SM technique.
\par
For simplicity, each CoMP BS is assumed to have two antennas $(T_{Xi}=2)$. $T_{Xi}$ represents the number of the $i^{th}$ BS antenna. Considering SM based space shift keying (SSK), it activates an antenna to transmit frequency for NOMA users while another antenna is utilized as an index modulation [15,16]. It is important to be noted that generalized SSK (GSSK) is required by putting more BS antennas while SM users require more capacity or $K$ number of users are considered [17,19].
\par
In this paper, both the first BS and the second BS coordinate to transmit information to one of the SM users (CoSM user), i.e. $CEU_k$ for $k=2$, by using the antenna index. In addition, the other SM users are called as non-coordinated SM (non-CoSM) users, i.e. $CEU_k$ for $3 \leq k \leq M$. Each $k^{th}$ non-CoSM gets information from the $i^{th}$ BS, where $i=k$, by using its antenna index. Table 1 explains that the number of antenna index combinations will increase if the first and the second BS perform CoSM. As a result of combination, CoSM user has four antenna combinations that each antenna may convey two input bits, while non-CoSM users only has two antenna combinations. It means each antenna of $i^{th}$ BS, for $3 \leq i \leq M$, only able to convey bit "0" or "1" delivered to non-CoSM users.
\par
In the receiver side, sometimes users cannot estimate information modulated through its antenna index perfectly. It may decrease user capacity, which is represented by a probability of error ($P_e$) [20-21]. This paper assumes $P_e$ only comes from the error estimation of binary phase-shift keying (BPSK) constellation symbol, while SM users are assumed may detect its antenna index perfectly.

\begin{table}[!b]
    \centering
    \caption {Mapping information of the antenna index.}
    \includegraphics [height=3.5cm,width=8.8cm]{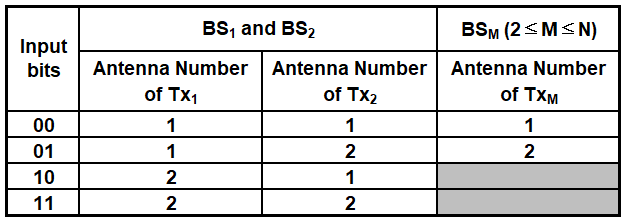}\\
    \label{Table1}
    \end{table}

\section{Ergodic Sum Capacity (ESC)} \label{sec_erg}
\subsection{ESC of JT-CoMP NOMA system}
In this section, ESC of JT-CoMP NOMA system is determined through a mathematical analysis based on Shannon formula. Considering $N$ number of cells, this paper analyzes ESC for $M$ number of CoMP BSs, while the other BSs do not apply coordination.
\par
By modifying received SINR in [7], then SINR of the $j^{th}$ CCU for $1 \leq j \leq M$ can be derived as
\begin{align*}
\tau _{j} &=\frac{\alpha \rho |h_{jj}|^2}{\underbrace{\alpha \rho \sum\limits_{\substack{i=1 \\ i \neq j}}^{M}|h_{ij}|^2 }_{\textit{ICI CoMP}}+ \underbrace{\rho \sum\limits_{\substack{i=M+1}}^{N}|h_{ij}|^2}_{\textit{ICI non-CoMP}}+ 1 }. \numberthis \label{eqn3.1_1}
\end{align*}
In equation (\ref{eqn3.1_1}), CCU suffers ICI from non-CoMP BSs, and ICI from CoMP BSs caused by power allocated to CCU. Considering JT-CoMP NOMA technique, CCU may cancel CoMP user signal by using SIC when the other BSs transmit the same data to CoMP user. Furthermore, imperfect CSI and imperfect SIC should be considered while users cannot estimate its channel and CCU cannot decode its signal perfectly. Then, SINR of the $j^{th}$ CCU for $1 \leq j \leq M$ can be derived as
\begin{align*}
\tau_{j} &=\frac{\alpha \rho |\hat{h}_{jj}|^2}{\alpha \rho \sum\limits_{\substack{i=1 \\ i \neq j}}^{M}|\hat{h}_{ij}|^2 + \rho \sum\limits_{\substack{i=M+1}}^{N}|\hat{h}_{ij}|^2+ \underbrace{\rho \sum\limits_{i=1}^{N} \sigma_{\epsilon_{ij}}}_{\textit{Im CSI}}     + \underbrace{\rho \gamma}_{\textit{Im SIC}} + 1 }. \numberthis \label{eqn3.1_2}
\end{align*}
The achievable data rate of the $j^{th}$ CCU for $1 \leq j \leq M$ can be written as
\begin{align*}
C_{j} &=log_2 \left( 1+ \tau_{j} \right). \numberthis \label{eqn3.1_3}\\
\end{align*}
\par
Similarly, by modifying received SINR for CoMP user in [7], SINR of the $k^{th}$ CEU for $k=1$ can be written as
\begin{align*}
\tau_{k}&=\frac{\beta\rho\sum\limits_{\substack{i=1}}^{M}\left | h_{ik} \right |^{2}}{\underbrace{\alpha\rho\sum\limits_{\substack{i=1}}^{M}\left | h_{ik} \right |^{2}}_{\textit{ICI CoMP}}+\underbrace{\rho \sum\limits_{\substack{i=M+1}}^{N}|h_{ij}|^2}_{\textit{ICI non-CoMP}}+1}.\numberthis \label{eqn3.1_4}
\end{align*}
Based on equation (\ref{eqn3.1_4}), the desired signal of CoMP user will increase if the number of CoMP BSs is increased, while it has the same interference pattern with CCU as shown in equation (\ref{eqn3.1_1}). Considering imperfect CSI, equation (\ref{eqn3.1_4}) can be written as
\begin{align*}
\tau_{k}&=\frac{\beta\rho\sum\limits_{\substack{i=1}}^{M}\left | \hat{h}_{ik} \right |^{2}}{\alpha\rho\sum\limits_{\substack{i=1}}^{M}\left | \hat{h}_{ik} \right |^{2}+\rho \sum\limits_{\substack{i=M+1}}^{N}|\hat{h}_{ij}|^2+\rho\sum\limits_{\substack{i=1}}^{N}\sigma_{\varepsilon{ik}}+1}. \numberthis \label{eqn3.1_5}
\end{align*}
The achievable data rate of the $k^{th}$ CEU for $k=1$ can be written as
\begin{align*}
C_{k}&=\log_{2}\left(1+\tau_k\right). \numberthis \label{eqn3.1_6}
\end{align*}

Moreover, ESC of JT-CoMP NOMA system can be expressed as follow
\begin{equation}
C_{\textit{JT-CoMP NOMA}}^{erg}=C_k+\sum_{j=1}^{M}C_{j}, \numberthis \label{eqn3.1_7}
\end{equation}
where $k=1$.
\par
The probability density function (PDF) of CCU capacity, i.e. $C_j^{exact}$, can be determined based on equation (\ref{eqn3.1_3}).
By using $\log_n(1+x/y)=\log_n\left[(x+y)/y \right]$ and $\log_n(x/y) = \log_n(x)-\log_n(y)$, then equation (\ref{eqn3.1_3}) can be simplified as%equation (\ref{eqn3.1_8}).
%
%\begin{figure*}
%   \hrulefill
   \begin{align*}
   C_{j}&
=log_2 \left( \frac{\alpha \rho \sum\limits_{i=1}^{M}|\hat{h}_{ij}|^2+\rho \sum\limits_{i=M+1}^{N}|\hat{h}_{ij}|^2 + \rho \sum\limits_{i=1}^{N} \sigma_{\varepsilon_{ij}}+\rho \Upsilon +1}{\alpha \rho \sum\limits_{\substack{i=1 \\ i \neq j}}^{M}|\hat{h}_{ij}|^2 + \rho \sum\limits_{i=M+1}^{N}|\hat{h}_{ij}|^2+ \rho \sum\limits_{i=1}^{N} \sigma_{\varepsilon_{ij}} + \rho \Upsilon + 1 }\right)\\
&=\log_2 \left(\alpha \rho \sum\limits_{i=1}^{M}|\hat{h}_{ij}|^2+\rho \sum\limits_{i=M+1}^{N}|\hat{h}_{ij}|^2 + \rho \sum\limits_{i=1}^{N} \sigma_{\varepsilon_{ij}}+\rho \Upsilon +1 \right) \\
&\ \ \ -
\log_2 \left(\alpha \rho \sum\limits_{\substack{i=1 \\ i \neq j}}^{M}|\hat{h}_{ij}|^2 + \rho \sum\limits_{i=M+1}^{N}|\hat{h}_{ij}|^2+ \rho \sum\limits_{i=1}^{N} \sigma_{\varepsilon_{ij}} + \rho \Upsilon + 1 \right).\numberthis \label{eqn3.1_8}
   \end{align*}
%   \hrulefill
%\end{figure*}
%
Furthermore, PDF of equation (\ref{eqn3.1_8}) for $C_j^{exact}$ is given by
\begin{align*}
C_{j}^{exact} &= E\{C_{j}\} \\
&= \int_{0}^{\infty} \log_2 \left( x + a \right) f_{X_j}(x)dx\\
&\ \ \ -\int_{0}^{\infty} \log_2 \left( y + a \right) f_{Y_j}(y) dy, \numberthis \label{eqn3.1_9}
\end{align*}
where $E(.)$ is denoted as the expectation operator and $a=\rho \sum_{i=1}^{N} \sigma_{\varepsilon_{ij}}+\rho\gamma+1$. By using equation (\ref{eqnA.1_3}) and (\ref{eqnA.1_4}), the PDF of equation (\ref{eqn3.1_9}) can be determined as
\begin{align*}
C_{j}^{\text{exact}} &= \int_{0}^{\infty} \log_2 \left( x + a \right) \sum\limits_{i=1}^{M} f_{X_{ij}}(x) \prod\limits_{\substack{h=1 \\ h \neq i}}^{N} \frac{k_{hj}}{k_{hj} - k_{ij}} dx \\
&\ \ \ -\int_{0}^{\infty} \log_2 \left( y + a \right) \sum\limits_{\substack{i=1 \\ i \neq j}}^{M}  f_{Y_{ij}}(y) \prod\limits_{\substack{h=1 \\ h \neq i \\ h \neq j}}^{N} \frac{k_{hj}}{k_{hj} - k_{ij}} dy. \numberthis \label{eqn3.1_10}
\end{align*}
By substituting equation (\ref{eqnA.1_1}) and equation  (\ref{eqnA.1_2}) into equation  (\ref{eqn3.1_10}), then $C_j^{exact}$ can be written as
\begin{align*}
C_{j}^{exact} &= \int_{0}^{\infty} \log_2 \left( x + a \right) \sum\limits_{i=1}^{M} k_{ij} \exp(-k_{ij} x)\\ 
&\ \ \ \ \times\prod\limits_{\substack{h=1 \\ h \neq i}}^{N} \frac{k_{hj}}{k_{hj} - k_{ij}} dx \\
&\ \ \ -\int_{0}^{\infty} \log_2 \left( y + a \right) \sum\limits_{\substack{i=1 \\ i \neq j}}^{M} k_{ij} \exp(-k_{ij} y)\\
&\ \ \ \ \times\prod\limits_{\substack{h=1 \\ h \neq i \\ h \neq j}}^{N} \frac{k_{hj}}{k_{hj} - k_{ij}} dy. \numberthis \label{eqn3.1_11}
\end{align*}

By using $\int_{0}^{\infty }exp(-Ax)ln(B+x)dx$=$\frac{1}{A}\left [ ln(B)exp(AB) \right ]$ and $log_2(x)=\frac{ln(x)}{ln(2)}$, then achievable data rate $C_j^{exact}$ can be written as
\begin{align*}
C_{j}^{exact}
&= \frac{1}{\ln(2)} \sum\limits _{i=1}^{M} \left( \ln(a) - \exp(ak_{ij}) \textrm{Ei}(-ak_{ij}) \right) \\
&\ \ \ \ \times \prod\limits_{\substack{h=1 \\ h \neq i}}^{N} \frac{k_{hj}}{k_{hj} - k_{ij}} \\
&\ \ \ -\frac{1}{\ln(2)} \sum\limits _{\substack {i=1 \\ i \neq j}}^{M} \left( \ln(a) - \exp(ak_{ij}) \textrm{Ei}(-ak_{ij}) \right) \\
&\ \ \ \ \times \prod\limits_{\substack{h=1 \\ h \neq i \\ h \neq j}}^{N} \frac{k_{hj}}{k_{hj} - k_{ij}}. \numberthis \label{eqn3.1_12}
\end{align*}

The PDF of equation (6) for $C_{k}^{exact}$ is given as
\begin{align*}
C_k^{exact} &= E\left\{ C_k \right\} \\
&= \int_{0}^{\infty} \log_2 \left( x + b \right) f_{X_k}(x)dx \\
&\ \ \ -\int_{0}^{\infty} \log_2 \left( y + b \right) f_{Y_k}(y) dy, \numberthis \label{eqn3.1_13}
\end{align*}
where $b=\rho \sum_{i=1}^{N} \sigma_{\varepsilon_{ik}}+1$. Similarly, by using equation (\ref{eqnA.1_1}), (\ref{eqnA.1_2}), (\ref{eqnA.1_3}) and (\ref{eqnA.1_4}), achievable data rate for $C_{k}^{\text{exact}}$ can be written as
\begin{align*}
C_{k}^{exact}
&= \frac{1}{\ln(2)} \sum\limits _{i=1}^{M} \left( \ln(b) - \exp(bl_{ik}) \textrm{Ei}(-bl_{ik}) \right) \\
&\ \ \ \ \times \prod\limits_{\substack{h=1 \\ h \neq i}}^{N} \frac{l_{hk}}{l_{hk} - l_{ik}} \\
&\ \ \ -\frac{1}{\ln(2)} \sum\limits _{\substack {i=1}}^{M} \left( \ln(b) - \exp(bm_{ik}) \textrm{Ei}(-bm_{ik}) \right) \\
&\ \ \ \ \times \prod\limits_{\substack{h=1 \\ h \neq i}}^{N} \frac{m_{hk}}{m_{hk} - m_{ik}}. \numberthis \label{eqn3.1_14}
\end{align*}
\par
Moreover, ESC of JT-CoMP NOMA system can be expressed as follow
\begin{equation}
C_{\textit{JT-CoMP NOMA}}^{exact}=C_{k}^{exact}+\sum_{j=1}^{M}C_{j}^{exact}, \numberthis \label{eqn3.1_15}
\end{equation}
where $k=1$.

\subsection{ESC of SM system}
In this section, ESC of SM system in $M$ number of CoMP BSs is determined based on the SSK scheme. Probability of error for each $k^{th}$ SM user, i.e. $P_{e,k}$ for $2 \leq k \leq M$, is calculated by comparing the number of error information received by SM users, and the number of original information transmitted from its BS. Considering Table 1, given $P_{e,k}$ in Appendix (A.3) and SSK capacity in [15], then achievable data rate for SM users can be determined as

%\begin{equation}
%C_{\hat{2}}=(1-P_{e,{\hat{2}}})\left \lfloor log_2(M_{T1}+M_{T2})  %\right \rfloor, & \mbox{for $N=2$}, \numberthis \label{eqn12}
%\end{equation}
%\begin{equation}
%C_k=(1-P_{e,k})\left \lfloor log_2(M_{Ti})  \right \rfloor, & %\mbox{for $N \geq 3$.}\numberthis \label{eqn13}
%\end{equation}

% rumus generalize
\begin{equation}
  C_k=\begin{cases}
    \left(1-P_{e,k}\right) \left \lfloor log_2 \left( \sum\limits_{i=1}^{2}T_{Xi} \right)  \right \rfloor &\mbox{, for $k= 2$}, \\
    \left(1-P_{e,k}\right) \left \lfloor log_2 \left( T_{X_k}\right)  \right \rfloor & \mbox{, for $3 \leq k \leq M$.}
  \end{cases} \numberthis \label{eqn3.2_1}
\end{equation}
Therefore, ESC for SM system can be written as 
\begin{equation}
C_{SM}^{erg}=\sum\limits _{k=2}^{M}C_k. \numberthis \label{eqn3.2_2}
\end{equation}
By substituting equation (\ref{eqnA.3_8}) and equation (\ref{eqnA.3_9}) into equation (\ref{eqn3.2_1}), then exact ESC of SM system can be assigned as 
\begin{equation}
C_{SM}^{exact}=\sum\limits _{k=2}^{M}C_k^{exact}. \numberthis \label{eqn3.2_3}
\end{equation}
\subsection{ESC of JT-CoMP NOMA-SM system}
Considering $N$ number of cells, this section calculates ESC for $M$ number of CoMP BSs, while the other BSs do not apply coordination as explained in Section \ref{sec_sys}. ESC of the proposed system can be determined by summing the total of both CCU and CEU capacity as follow

\begin{align*}
C_{\textit{JT-CoMP NOMA-SM}}^{erg}&=\sum_{j=1}^{M}C_{j}+\sum_{k=1}^{M}C_{k}\\
&=\left(\sum_{j=1}^{M}C_{j}+C_{k=1}\right)+\sum_{k=2}^{M}C_{k}\\
&=C_{\textit{JT-CoMP NOMA}}^{erg}+C_{SM}^{erg}. \numberthis \label{eqn3.3_1}
\end{align*}
Similarly, exact ESC of the proposed system can be written as
\begin{equation}
C_{\textit{JT-CoMP NOMA-SM}}^{exact}=C_{\textit{JT-CoMP NOMA}}^{exact}+C_{SM}^{exact}. \numberthis \label{eqn3.3_2}
\end{equation}
\par
%Furthermore, equation (\ref{eqn3.3_1}) and equation (\ref{eqn3.3_2}) can be divided by the number of cells to calculate average ESC and exact average ESC of the proposed system, respectively. %Therefore, average ESC of the proposed system can be calculated as
%
%\begin{align*}
%\bar{C}_{\textit{JT-CoMP NOMA-SM}}^{erg}&=\frac{C_{\textit{JT-CoMP NOMA-SM}}^{erg}}{N}
%.\numberthis \label{eqn3.3_3}
%\end{align*}
%Moreover, exact average ESC of the proposed system can be written as
%\begin{equation}
%\bar{C}_{\textit{JT-CoMP NOMA-SM}}^{exact}=\frac{C_{\textit{JT-CoMP NOMA}}^{exact}+C_{SM}^{exact}}{N}. \numberthis \label{eqn3.3_4}
%\end{equation}
%
\section{Simulation Result and Analysis} \label{sec_result}

This section analyzes performances of the proposed system such as CCU capacity, CEU capacity, and ESC compared to NOMA and JT-COMP VP-NOMA. For simplicity, these are simulated and analyzed based on a simplified three-CoMP BS ($M=3$) within a twelve-cell model ($N=12$).
\begin{figure}[!t]
    \centering
    \includegraphics [width=0.6\textwidth]{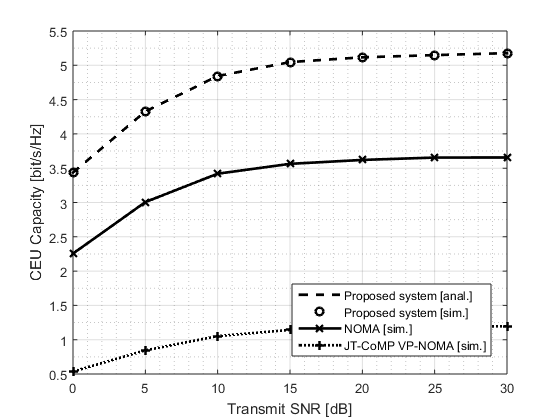} \\
    \caption {CEU capacity with respect to transmit SNR ($\rho$); $M=3$, $N=12$, $\sigma_{\varepsilon ik} = 0.01$, and $\gamma = -25$ dB.}
    \label{Fig2}
    \end{figure}
\par
\begin{figure}[!b]
    \centering
    \includegraphics [width=0.6\textwidth]{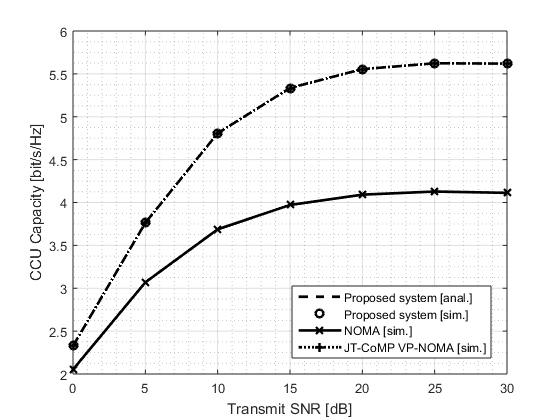} \\
    \caption {CCU capacity with respect to transmit SNR ($\rho$); $M=3$, $N=12$, $\sigma_{\varepsilon ij} = 0.01$, and $\gamma = -25$ dB.}
    \label{Fig3}
    \end{figure}

Fig. \ref{Fig2} shows that CEU capacity of the proposed system is the highest than the other schemes. CEU capacity of the proposed system increases by 41.3\% compared to NOMA at $\rho=20$ dB. It is interesting that JT-CoMP VP-NOMA shows the worst CEU capacity. It happens because JT-CoMP VP-NOMA divides its power to one CCU and three CEUs ($M$=3), while the proposed system uses conventional CoMP and SM simultaneously can optimize its bandwidth and power. On the other hand, the proposed system has the same CCU capacity compared to JT-CoMP VP-NOMA as shown in Fig. \ref{Fig3}. It is caused CCU of the proposed system has the same interference pattern as JT-CoMP VP-NOMA. By using JT-CoMP technique of both two schemes compared, CCU can mitigate most of CoMP user signals by using SIC while imperfect SIC is considered. 

\begin{figure}[!t]
    \centering
    \includegraphics [width=0.6\textwidth]{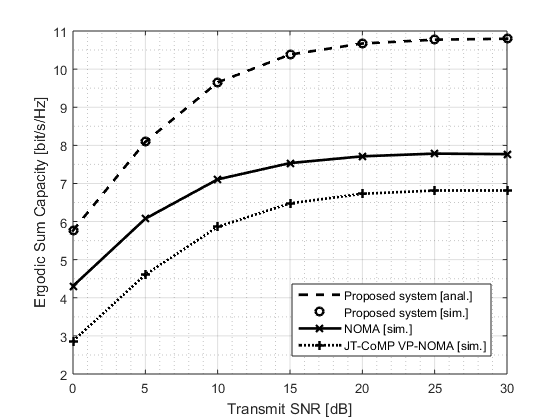}\\
    \caption {ESC with respect to transmit SNR ($\rho$); $M=3$, $N=12$, $\sigma_{\varepsilon} = 0.01$, and $\gamma = -25$ dB.}
    \label{Fig4}
    \end{figure}
\begin{figure}[!b]
    \centering
    \includegraphics [width=0.6\textwidth]{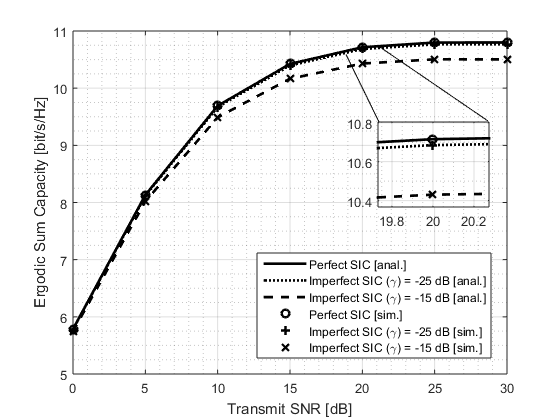}\\
    \caption {ESC of the proposed system with respect to transmit SNR ($\rho$); $M=3$, $N=12$, and $\sigma_{\varepsilon} = 0.01$.}
    \label{Fig5}
    \end{figure}
In this paper, each variance of the error estimation parameter $\sigma_{\varepsilon ij}$ and $\sigma_{\varepsilon ik}$ is assumed to have the same value expressed by $\sigma_{\varepsilon}$. Considering $\sigma_{\varepsilon}=0.01$, Fig. \ref{Fig4} shows that the proposed system has 38.3\% and 58.5\% higher ESC at $\rho=20$ dB than NOMA and JT-CoMP VP-NOMA, respectively. Based on Fig. \ref{Fig2} and Fig. \ref{Fig3}, it is clear that increasing ESC of the proposed system is caused due to enhancing CEU capacity, while CCU capacity can be maintained. 

\begin{figure}[!t]
    \centering
    \includegraphics [width=0.6\textwidth]{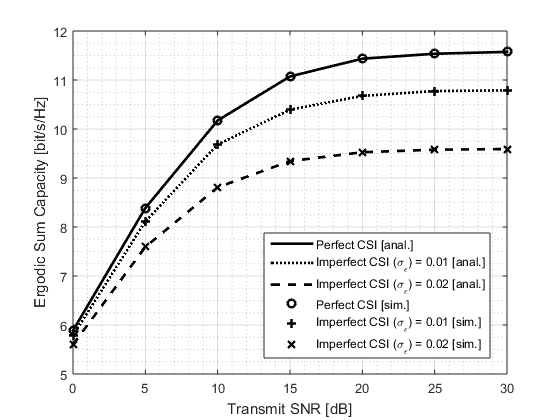}\\
    \caption {ESC of the proposed system with respect to transmit SNR ($\rho$); $M=3$, $N=12$, and $\gamma = -25$ dB.}
    \label{Fig6}
    \end{figure}
\begin{figure}[!b]
    \centering
    \includegraphics [width=0.6\textwidth]{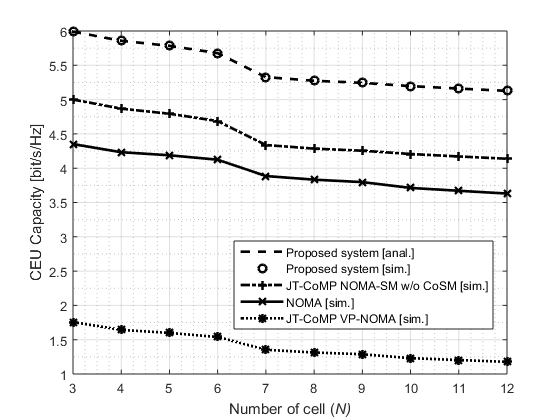}\\
    \caption {CEU capacity with respect to the number of cells ($N$); $M=3$, $\rho = 20$ dB, $\sigma_{\varepsilon} = 0.01$, and $\gamma = -25$ dB.}
    \label{Fig7}
    \end{figure}
\par
Fig. \ref{Fig5} shows that considering $\gamma=-15$ dB and $\gamma=-25$ dB can degrade 2.7\% and 0.28\% ESC of the proposed system at $\rho=20$ dB respectively, compared to perfect SIC. On the other hand, by considering $\sigma_{\varepsilon}=0.01$ and $\sigma_{\varepsilon}=0.02$, it may degrade 6.56\% and 16.65\% ESC of the proposed system respectively at $\rho=20$ dB compared to perfect SIC as shown in Fig. \ref{Fig6}. Furthermore, Fig. \ref{Fig5} and Fig. \ref{Fig6} prove that both imperfect SIC and imperfect CSI can lead to additional interference may degrade capacity.
%
%\begin{figure}[!b]
%    \centering
%    \includegraphics [width=0.45\textwidth]{Fig7.png}\\
%    \caption {Average ESC with respect to transmit SNR ($\rho$); $N=12$, $\sigma_{\varepsilon} = 0.01$, and $\gamma = -25$ dB.}
%    \label{Fig7}
%   \end{figure}
% 
\par
%In Fig. \ref{Fig7}, average ESC of both the proposed system and JT-CoMP VP-NOMA decrease if the number of CoMP BSs is increased. In the proposed system, CEU located in the third cell cannot utilize CoSM because of the limited number of SM antenna used. In JT-CoMP VP-NOMA, both bandwidth and BS power should be divided by one CCU and three CEUs ($M=3$) if the number of CoMP BSs is increased. In this case, implementing two CoMP BSs ($M=2$) within twelve cells ($N=12$) model can achieve higher average ESC while the other BSs do not apply coordination.
\par
For $N=3$ to $N=12$, CEU capacity of the proposed system decreases 14.42\% while NOMA and JT-CoMP VP-NOMA degrade 16.57\% and 32.93\% respectively as shown in Fig. \ref{Fig7}. It proves that the proposed system can maintain CEU capacity better than the other schemes if the number of cells is increased. It is caused SM users do not suffer ICI because SM works beyond Shannon upper bounds. In addition, utilizing CoSM may improve around 1 bit/s/Hz higher CEU capacity than without using CoSM.

\section{Conclusion} \label{sec_conclusion}
In this paper, JT-CoMP NOMA-SM is proposed to enhance capacity. The results show that proposed system has the same CCU capacity compared to JT-CoMP VP-NOMA. Considering $\rho=20$ dB, the proposed system has 41.3\% higher CEU capacity compared to NOMA. Furthermore, the proposed system has 38.3\% and 58.5\% higher ESC than NOMA and JT-CoMP VP-NOMA, respectively. Imperfect SIC and imperfect CSI  may degrade capacity. %Average ESC of the proposed system decreases if the number of CoMP BSs is increased. 
The proposed system can maintain CEU capacity better than the other schemes if the number of cells is increased. By utilizing CoSM, CEU capacity increases around 1 bit/s/Hz higher than without using CoSM. Moreover, the proposed system outperforms the other schemes. 
\par
For future work, a different method of ICI avoidance and SM techniques can be exploited to achieve higher spectral efficiency.

%
%\section*{Acknowledgment}
%This work was supported by Brain Korea 21 Plus Project (Dept. of IT Convergence %Engineering, Kumoh National Institute of Technology).
%

\section*{Appendix A}
\subsection*{A.1. Derivation of $f_{X_j}(x)$ and $f_{Y_j}(y)$ for CCU capacity}
Let suppose $X_j$$\triangleq$$\alpha \rho \sum_{i=1}^{M}|\hat{h}_{ij}|^2+\rho \sum\limits_{i=M+1}^{N}|\hat{h}_{ij}|^2$ and $Y_j$$\triangleq$$\alpha \rho \sum_{i=1,i \neq j}^{M}|\hat{h}_{ij}|^2+\rho \sum\limits_{i=M+1}^{N}|\hat{h}_{ij}|^2$. Considering each $j$ symbol mentioned in this section represents the $j^{th}$ CCU for $1 \leq j \leq M$. then PDF of $X_j$ can be determined by using the sum of the number of cells. In this case, random variables are assumed independent and identically distributed exponential with different parameters. In addition, $Y_j$ can be determined by using the sum of the $i^{th}$ cell, where $i \neq j$. The parameters of exponential random variables are assumed to be different because the distance from each $i^{th}$ BS to the $j^{th}$ CCU is different. The PDF for each exponential random variable can be determined as
\begin{align*}
f_{X_{ij}}(x) &= \frac{d(F_{X_{ij}}(x))}{dx} = \frac{d(1- \exp(-k_{ij} x))}{dx} \\
&= k_{ij} \exp(-k_{ij} x), \numberthis \label{eqnA.1_1}
\end{align*}
\begin{align*}
f_{Y_{ij}}(y) &= \frac{d(F_{Y_{ij}}(y))}{dy} = \frac{d(1- \exp(-k_{ij} y))}{dy} \\
&= k_{ij} \exp(-k_{ij} y),\numberthis \label{eqnA.1_2}
\end{align*}
where $k_{ij}$ represents each parameter of exponential random variables. Then, the PDF of $X_j$ and $Y_j$ for $1 \leq j \leq M$ within $N$ number of cells can be written as 
\begin{align*}
f_{X_{j}}(x) &= f_{X_{ij}+...+X_{Mj}+...+X_{Nj}}(x)  \\
&=  \sum\limits_{i=1}^{M} f_{X_{ij}}(x) \prod\limits_{\substack{h=1 \\ h \neq i}}^{N} \frac{k_{hj}}{k_{hj} - k_{ij}}, & \mbox{$2 \leq M \leq N$}, \numberthis \label{eqnA.1_3}
\end{align*}
\begin{align*}
f_{Y_{j}}(y) &= f_{Y_{ij}+...+Y_{Mj}+...+Y_{Nj}}(y), & \mbox{$i \neq j$}, \\
&=  \sum\limits_{\substack{i=1 \\ i \neq j}}^{M}  f_{Y_{ij}}(y) \prod\limits_{\substack{h=1 \\ h \neq i \\ h \neq j}}^{N} \frac{k_{hj}}{k_{hj} - k_{ij}}, & \mbox{$3 \leq M \leq N$}.\numberthis \label{eqnA.1_4}
\end{align*}
Considering $M$ number of CoMP BSs within $N$ number of cells, then $k_{ij}$ can be assigned as
\begin{align*}
  k_{ij}=\begin{cases}
    \frac{1}{\alpha\rho\hat{\sigma}_{ij}}, & \mbox{for $1 \leq i \leq M$,}\\
    \frac{1}{(\alpha+\beta)\rho\hat{\sigma}_{ij}}, & \mbox{for $M+1 \leq i \leq N$.}
  \end{cases} \numberthis \label{eqnA.1_5}
\end{align*}

\subsection*{A.2. Derivation of $f_{X_k}(x)$ and $f_{Y_k}(y)$ for CEU capacity}
Let suppose $X_{k}$$\triangleq$$\rho \sum_{i=1}^{N}|\hat{h}_{ik}|^2$ and $Y_{k}$$\triangleq$$\alpha \rho \sum_{i=1}^{M}|\hat{h}_{ik}|^2+\rho \sum\limits_{i=M+1}^{N}|\hat{h}_{ik}|^2$. In this section, each $k$ symbol mentioned represents the $k^{th}$ CEU for $k=1$. Similarly, PDF for each exponential random variable can be determined as
\begin{align*}
f_{X_{ik}}(x) &= \frac{d(F_{X_{ik}}(x))}{dx} = \frac{d(1- \exp(-l_{ik} x))}{dx} \\
&= l_{ik} \exp(-l_{ik} x), \numberthis \label{eqnA.2_1}
\end{align*}
\begin{align*}
f_{Y_{ik}}(y) &= \frac{d(F_{Y_{ik}}(y))}{dy} = \frac{d(1- \exp(-m_{ik} y))}{dy} \\
&= m_{ik} \exp(-m_{ik} y),\numberthis \label{eqnA.2_2}
\end{align*}
where $l_{ik}=\frac{1}{(\alpha+\beta)\rho\hat{\sigma}_{ik}}$. In addition, $m_{ik}$ depends on the number of CoMP BSs. Then, PDF of $X_{k}$ and $Y_{k}$ for $k=1$ can be written as (\ref{eqnA.2_3}) and (\ref{eqnA.2_4}), respectively.
\begin{align*}
f_{X_{k}}(x) &= f_{X_{ik}+...+X_{Mk}+...+X_{Nk}}(x)  \\
&=  \sum\limits_{i=1}^{M} f_{X_{ik}}(x) \prod\limits_{\substack{h=1 \\ h \neq i}}^{N} \frac{l_{hk}}{l_{hk} - l_{ik}}, & \mbox{$2 \leq M \leq N$},\numberthis \label{eqnA.2_3}
\end{align*}
\begin{align*}
f_{Y_{k}}(y) &= f_{Y_{ik}+...+Y_{Mk}+...+Y_{Nk}}(y)  \\
&=  \sum\limits_{i=1}^{M} f_{Y_{ik}}(y) \prod\limits_{\substack{h=1 \\ h \neq i}}^{N} \frac{m_{hk}}{m_{hk} - m_{ik}}, & \mbox{$2 \leq M \leq N$}.\numberthis \label{eqnA.2_4}
\end{align*}

Considering $M$ number of CoMP BSs within $N$ number of cells, then $m_{ik}$ can be assigned as 
\begin{align*}
  m_{ik}=\begin{cases}
    \frac{1}{\alpha\rho\hat{\sigma}_{ik}}, & \mbox{for $1 \leq i \leq M$,}\\
    \frac{1}{(\alpha+\beta)\rho\hat{\sigma}_{ik}}, & \mbox{for $M+1 \leq i \leq N$.}
  \end{cases} \numberthis \label{eqnA.2_5}
\end{align*}
\subsection*{A.3. Derivation probability of error} \label{appendix}
In this case, BS transmits a signal for SM users mapped in BPSK constellation symbol. Considering information of SM users is superimposed through power $P$, then the transmitted signal is given as
\begin{align*}
x_k &=\sqrt{P} \ S_{k}, &\mbox{for $2 \leq k \leq M$},  \numberthis \label{eqnA.3_1}
\end{align*}
where $S_k$ represents coded signals of the $k^{th}$ SM user. Considering $BS_1$ and $BS_2$ make coordination each other to transmit information to CoSM user, then received signal for the $k^{th}$ CoSM user, i.e. $y_k$ for $k=2$, is given as
\begin{align*}
y_{k}&=y_{1{k}}^{*}+y_{2{k}}^{*}\\
&= \left[\hat{h}_{1{k}}\sqrt{P}S_{k}+n_{1{k}} \right]+\left[\hat{h}_{2{k}}\sqrt{P}S_{k}+n_{2{k}} \right]\\
&=(\hat{h}_{1{k}}+\hat{h}_{2{k}})\sqrt{P}S_{k}+n_{k},  \numberthis \label{eqnA.3_2}
\end{align*}
Furthermore, the other BSs ($BS_3,...,BS_M$) transmit information symbols to non-CoSM users ($CEU_3,...,CEU_M$) independently. Therefore, the received signal for each $k^{th}$ non-CoSM user, i.e. $y_k$ for $3 \leq k \leq M$, is given as
\begin{align*}
y_{k}&=\hat{h}_{ik}\sqrt{P}S_{k}+n_{k}, &\mbox{$i=k$}. \numberthis \label{eqnA.3_3}
\end{align*}
%Furthermore, the SM users decode own information as follows

%\begin{equation}
%\mu_{\hat{2}}^{*}=pskdemod\left(\frac{y_{1{\hat{2}}}^{*}}{\hat{h}_{1{%\hat{2}}}},M\right)+pskdemod\left(\frac{y_{2{\hat{2}}}^{*}}{\hat{h}_{%1{\hat{2}}}},M\right), \numberthis \label{eqn31}
%\end{equation}
%\begin{equation}
%\mu_{\hat{k}}^{*}=pskdemod\left(\frac{y_{{\hat{k}}}}{\hat{h}_{3{\hat{%k}}}},M\right). \numberthis \label{eqn32}
%\end{equation}

%\begin{equation}
%  \mu_{\hat{k}}^{*}=\begin{cases}
%   pskdemod\left(\frac{y_{1{\hat{2}}}^{*}}{\hat{h}_{1{\hat{2}}}},M\right)+%pskdemod\left(\frac{y_{2{\hat{2}}}^{*}}{\hat{h}_{1{\hat{2}}}},M\right), %\mbox{$N \leq 2$}, \\
%    pskdemod\left(\frac{y_{{\hat{k}}}}{\hat{h}_{N{\hat{k}}}},M\right),\ \ %\ \ \ \ \ \ \ \ \ \ \ \ \ \ \ \ \ \ \ \ \ \ \ \ \ \ \ \mbox{$N \geq 3$}.
%  \end{cases} \numberthis \label{eqn31}
%\end{equation}

Considering $\hat{y}_{k}$ represents the desired signal at SM users, decoded information of $\hat{y}_{k}$ may be different with $y_{k}$ due to error transmission. Given decoded signal in [21], then error detection for each $k^{th}$ SM user can be written as
\begin{align*}
 P_{e,{k}} &= \left \lfloor \frac{\hat{y}_{k}-y_{k}} {\sqrt{P}}  \right \rfloor, &\mbox{$2 \leq k \leq M$}. \numberthis \label{eqnA.3_4}
\end{align*}

The PDF of error probability for CoSM user, i.e. $P_{e,k}^{exact}$ for $k = 2$, can be calculated by solving
\begin{align*}
 P_{e,{k}}^{exact}=& \frac{1}{2} \left[\int_{0}^{\infty} Q(\sqrt{\delta_{1k}^*})f_{\delta_{1k}^*}(\delta_{1k}^*)d\delta_{1k}^*\right.\\
 &\left.+\int_{0}^{\infty}Q(\sqrt{\delta_{2k}^*})f_{\delta_{2k}^*}(\delta_{2k}^*)d\delta_{2k}^*\right],
 \numberthis \label{eqnA.3_5}
\end{align*}
where $\delta_{1k}^*=\rho\hat{h}_{1k}$ and $\delta_{2k}^*=\rho\hat{h}_{2k}$. In case each random variable of Rayleigh fading channel is distributed independently, then PDF of $\delta_{ik}^*$ can be derived as
\begin{align*}
f_{\delta_{ik}^*}(\delta)&=\frac{1}{\hat{\delta}_{ik}^*} exp(-\frac{\delta}{\hat{\delta}_{ik}^*}), &\mbox{$\delta \geq 0$},
 \numberthis \label{eqnA.3_6}
\end{align*}
where $\hat{\delta}_{ik}^*=\rho Ei\left(\hat{\sigma}_{ik}\right)$, for $k=2$ and $1 \leq i \leq 2$.
By using an alternative representation of $Q$ function defined in [22], the probability of error for CoSM user is given as
\begin{align*}
 P_{e,{k}}^{exact}=& \frac{1}{2} \left[\frac{1}{\pi}\int_{-\pi/2}^{\pi/2} MGF_{\delta_{1k}^*}\left(\frac{1}{sin^2\theta } \right)d\theta \right.\\
 &\left.+\frac{1}{\pi}\int_{-\pi/2}^{\pi/2}MGF_{\delta_{2k}^*}\left(\frac{1}{sin^2\theta } \right)d\theta\right],
 \numberthis \label{eqnA.3_7}
\end{align*}
where $MGF_{\delta_{ik}^*}(s)=\frac{1}{1+s\hat{\delta}_{ik}^*}$, for $k=2$ and $1 \leq i \leq 2$.
By modifying probability of error in [23], then $P_{e,k}^{exact}$ for $k=2$ is given as
\begin{align*}
 P_{e,{k}}^{exact}=& \frac{1}{2} \left[\left(1-\sqrt{\frac{\hat{\delta}_{1k}^*}{2+\hat{\delta}_{1k}^*}}\right)+\left(1-\sqrt{\frac{\hat{\delta}_{2k}^*}{2+\hat{\delta}_{2k}^*}}\right)\right].
 \numberthis \label{eqnA.3_8}
\end{align*}
\par
Similarly, the exact probability of error for each $k^{th}$ non-CoSM user, i.e. $P_{e,k}^{exact}$ for $3 \leq k \leq M$, can be written as
\begin{align*}
 P_{e,{k}}^{exact}=& \frac{1}{2} \left[\left(1-\sqrt{\frac{\hat{\delta}_{ik}}{2+\hat{\delta}_{ik}}}\right)\right],
 \numberthis \label{eqnA.3_9}
\end{align*}
where $\hat{\delta}_{ik}=\rho Ei\left(\hat{\sigma}_{ik}\right)$ for $3 \leq k \leq M$ and $i=k$.

%% References


%\bibliographystyle{elsarticle-num}
% \bibliographystyle{elsarticle-harv}
% \bibliographystyle{elsarticle-num-names}
% \bibliographystyle{model1a-num-names}
% \bibliographystyle{model1b-num-names}
% \bibliographystyle{model1c-num-names}
% \bibliographystyle{model2-names}
% \bibliographystyle{model3a-num-names}
% \bibliographystyle{model3-num-names}
% \bibliographystyle{model4-names}
% \bibliographystyle{model5-names}
% \bibliographystyle{model6-num-names}

%\bibliography{sample}

\section*{References}

\noindent [1] Z. Ding, X. Lei, G. K. Karagiannidis, R. Schober, J. Yuan, V. K. Bhargava, A survey on non-orthogonal multiple access for 5G networks: Research challenges and future trends, IEEE Journal on Selected Areas in Communications 35 (10) (2017) 2181-2195. doi:10.1109/JSAC.2017.2725519.\\
URL http://geokarag.webpages.auth.gr/wp-content/papercite-data/pdf/j230.pdf\\

\noindent [2] A. Benjebbour, A. L. K. Saito, Y. Kishiyama, T. Nakamura, Signal Processing for 5G: Algorithms and
Implementations, First Edition, John Wiley and Sons, Ltd, 2016. doi:10.1002/9781119116493.\\
URL http://solutionsproj.net/software/Signal\underline{ }Processing\underline{ }for\underline{ }5G\underline{ }Algorithms\underline{ }and\underline{ }
Implementations.pdf\\

\noindent [3] M. B. Shahab, M. F. Kader, S. Y. Shin, On the power allocation of non-orthogonal multiple access for 5G wireless networks, in: 2016 International Conference on Open Source Systems Technologies
(ICOSST), 2016, pp. 89-94. doi:10.1109/ICOSST.2016.7838583.\\
URL https://ieeexplore.ieee.org/stamp/stamp.jsp?tp=\&arnumber=7838583\\

\noindent [4] Y. Saito, Y. Kishiyama, A. Benjebbour, T. Nakamura, A. Li, K. Higuchi, Non-orthogonal multiple access (NOMA) for cellular future radio access, in: 2013 IEEE 77th Vehicular Technology Conference (VTC Spring), 2013, pp. 1-5. doi:10.1109/VTCSpring.2013.6692652.\\
URL https://ieeexplore.ieee.org/stamp/stamp.jsp?tp=\&arnumber=6692652\\

\noindent [5] J. Choi, Non-orthogonal multiple access in downlink coordinated two-point systems, IEEE Communications Letters 18 (2) (2014) 313-316. doi:10.1109/LCOMM.2013.123113.132450.\\
URL https://ieeexplore.ieee.org/stamp/stamp.jsp?arnumber=6708131\\

\noindent [6] Y. Sun, Z. Ding, X. Dai, G. K. Karagiannidis, A novel network NOMA scheme for downlink coordinated three-point systems, arXiv:1708.06498 [cs.IT], [Online]. https://arxiv.org/pdf/1708.06498.pdf.\\

\noindent [7] F. W. Murti, R. F. Siregar, S. Y. Shin, Exploiting non-orthogonal multiple access in downlink coordinated multipoint transmission with the presence of imperfect channel state information, Computer Sciences.\\
URL https://arxiv.org/abs/1812.10266\\

\noindent [8] M. S. Ali, E. Hossain, A. Al-Dweik, D. I. Kim, Downlink power allocation for CoMP-NOMA in multi-cell networks, IEEE Transactions on Communications 66 (9) (2018) 3982-3998. doi:10.1109/TCOMM.2018. 2831206.\\
URL https://ieeexplore.ieee.org/stamp/stamp.jsp?arnumber=8352643\\

\noindent [9] M. S. Ali, E. Hossain, D. I. Kim, Coordinated multipoint transmission in downlink multi-cell NOMA systems: Models and spectral efficiency performance, IEEE Wireless Communications 25 (2) (2018) 24-31. doi:10.1109/MWC.2018.1700094.\\
URL https://ieeexplore.ieee.org/stamp/stamp.jsp?tp=\&arnumber=8352618\\

\noindent [10] Y. Al-Eryani, E. Hossain, D. I. Kim, Generalized coordinated multipoint (GCoMP)-enabled NOMA: Outage, capacity, and power allocation, arXiv:1901.10535 [cs.IT], [Online]. https://arxiv.org/pdf/1901.10535.pdf.\\

\noindent [11] D. K. Hendraningrat, M. B. Shahab, S. Y. Shin, Virtual user pairing non-orthogonal multiple access in downlink coordinated multipoint transmission, arXiv:1903.10674 [cs.NI], [Online]. https://arxiv.org/abs/ 1903.10674.\\

\noindent [12] M. B. Shahab, S. Y. Shin, On the performance of a virtual user pairing scheme to efficiently utilize the spectrum of unpaired users in NOMA, Physical Communications 25 (2017) 492-501.\\
URL https://www.sciencedirect.com/science/article/pii/S1874490717302082\\

\noindent [13] E. Soujeri, Advanced Index Modulation Techniques for Future Wireless Networks, Ecole De Technologie Superieure, Universite Du Quebec, 2018.\\
URL http://espace.etsmtl.ca/2091/1/SOUJERI\underline{ }Ebrahim.pdf\\

\noindent [14] R. Y. Mesleh, H. Haas, S. Sinanovic, C. W. Ahn, S. Yun, Spatial modulation, IEEE Transactions on Vehicular Technology 57 (4) (2008) 2228-2241. doi:10.1109/TVT.2007.912136.\\
URL https://ieeexplore.ieee.org/stamp/stamp.jsp?arnumber=4382913\\

\noindent [15] J. W. Kim, M. Irfan, S. M. AL, S. Y. Shin, Selective non-orthogonal multiple access (NOMA) and spatial modulation (SM) for improved spectral efficiency, 2015 International Symposium on Intelligent Signal Processing and Communication Systems (ISPACS) (2015) 552-555. doi:10.1109/ISPACS.2015.7432833.\\
URL https://ieeexplore.ieee.org/stamp/stamp.jsp?tp=\&arnumber=7432833\\

\noindent [16] F. Kara, H. Kaya, Performance analysis of SSK-NOMA, arXiv:1905.00777 [cs.IT], [Online]. https://arxiv. org/pdf/1905.00777.pdf.\\

\noindent [17] M. Irfan, B. S. Kim, S. Y. Shin, A spectral efficient spatially modulated non-orthogonal multiple access for 5G, in: 2015 International Symposium on Intelligent Signal Processing and Communication Systems (ISPACS), 2015, pp. 625-628. doi:10.1109/ISPACS.2015. 7432847.\\
URL https://ieeexplore.ieee.org/stamp/stamp.jsp?tp=\&arnumber=7432847\\

\noindent [18] J. W. Kim, S. Y. Shin, V. C. M. Leung, Performance enhancement of downlink NOMA by combination with GSSK, IEEE Wireless Communications Letters 7 (5) (2018) 860-863. doi:10.1109/LWC.2018.2833469.\\
URL https://ieeexplore.ieee.org/stamp/stamp.jsp?tp=\&arnumber=8355593\\

\noindent [19] K. Ntontin, M. D. Renzo, A. Perez-Neira, C. Verikoukis, Adaptive generalized space shift keying, EURASIP Journal on Wireless Communications and Networking 2013 (1) (2013) 43. doi:10.13067/JKIECS. 2015.10.11.1257.\\
URL https://jwcn-eurasipjournals.springeropen.com/track/pdf/10.1186/1687-1499-2013-43\\

\noindent [20] R. F. Siregar, F. W. Murti, S. Y. Shin, Bit allocation approach of spatial modulation for multi-user scenario, Journal of Network and Computer Applications 127 (2019) 1-8. doi:https://doi.org/10.1016/j.jnca. 2018.11.014.\\
URL http://www.sciencedirect.com/science/article/pii/S1084804518303801\\

\noindent [21] M. R. Usman, A. Khan, M. A. Usman, Y. S. Jang, S. Y. Shin, On the performance of perfect and imperfect SIC in downlink non-orthogonal multiple access (NOMA), in: 2016 International Conference on Smart Green Technology in Electrical and Information Systems (ICSGTEIS), 2016, pp. 102-106. doi:10.1109/ICSGTEIS.2016.7885774.\\
URL https://ieeexplore.ieee.org/document/7885774\\

\noindent [22] J. W. Craig, A new, simple and exact result for calculating the probability of error for two-dimensional signal constellations, in: MILCOM 91 - Conference record, 1991, pp. 571-575 vol.2. doi:10.1109/MILCOM. 1991.258319.\\
URL https://ieeexplore.ieee.org/stamp/stamp.jsp?tp=\&arnumber=258319\\

\noindent [23] F. Kara, H. Kaya, BER performances of downlink and uplink NOMA in the presence of SIC errors over fading channels, IET Communications 12 (15) (2018) 1834-1844. doi:10.1049/iet-com.2018.5278.\\
URL https://ieeexplore.ieee.org/stamp/stamp.jsp?tp=\&arnumber=8457925

\end{document}